\documentclass[aps,pra, onecolumn,14 pts,a4paper,showpacs]{revtex4}
\usepackage{amssymb,amsmath,amsfonts,graphicx}
\usepackage{bm}

\begin{document}

\title{ A non-equilibrium system in a steady state: wind waves in the open ocean}

\author{ Yves Pomeau }
\affiliation{Department of Mathematics, University of Arizona,
Tucson, USA.}
\author{Martine Le Berre}
\affiliation{Institut des Sciences Moleculaires d'Orsay, Bat.210,
91405 Orsay, France.}
\date{\today }

\begin{abstract}
We derive scaling laws for the steady spectrum of wind excited waves, assuming two inviscid fluids (air and water)
and no surface tension, an approximation valid at large speeds. In this limit there exists an unique  (small) dimensionless parameter $\epsilon$
, the ratio of the mass
densities of the two fluids, air and water , independently of the wind speed.
 The smallness of $\epsilon$ allows to derive some important average properties of the wave system.
  The average square slope of the waves is of order $|ln (\epsilon^2)|^{-1}$,
  a small but not very small
  quantity. This supports the often used assumption of small nonlinearity in the wave-wave interaction. We introduce an
  equation to be satisfied by the two-point correlation of the height fluctuations.
\end{abstract}

\maketitle


\section{Introduction}
\label{sec:Intro}
Wave turbulence makes a problem interesting both from the point of view of fundamental science and because it has obvious
 connections with real life phenomena which play a significant role in many human endeavors. Wind-excited waves provide
  an example of a non-equilibrium system which may be in a steady
  state.
  After many scientists, we
 consider below the following
  problem: a constant wind blowing on the horizontal surface of an infinite Ocean (in the three dimensions), and
  exciting, after a transient, a system of
  random fluctuating waves that reach a turbulent steady state. After Pierson and Moskowitz \cite{PM}
  this makes what is called a "fully developed sea" in the oceanographic
  literature.
  One studies the statistical
  properties like the variance of
  the surface slope as well as
  the strength of the impact of the waves on various
structures and finally the power spectrum of the surface
elevation. This has been thought about for many years
  \cite{gelci}, \cite{hasselman}, mainly by using the weak interaction approximation, based on the (unexplained to the best of our knowledge) observation that
  the slope of waves is small on average.
  Below we approach this problem by using simple scaling ideas based on the existence of a small parameter, the ratio of
   the mass density of the air $\rho_{air}$ to the mass density of water $\rho_{w}$. Let
   $\epsilon = \frac{\rho_{air}}{\rho_{w}}$ be this small ratio, about $10^{-3}$. Assuming incompressibility,
   (and neglecting, for the moment, surface tension and viscosity), the data with a
physical dimension
   are the acceleration of gravity $g$ and the (uniform) wind speed $U$. As
   shown by Newton, one can make out of these two quantities  a length, $\lambda = \frac{U^2}{g}$ and a time, $\tau = \frac{\lambda}{U} = \frac{U}{g}$. Here we
   shall use $\lambda$ and $\tau$ as units of space and time, that leads
    to a system of equations without the physical parameters $g, U$ \cite {YPscaling}.
    This is in contrast with "ordinary" turbulence where length scales are fixed in two ways: the large scale is
    determined by the geometry of the flow, the diameter of a pipe for instance, although the small scale is the
     Kolmogorov scale where viscosity becomes important. Moreover we shall argue that, in wave turbulence, there is no step by
     step cascade from large to small scale where dissipation takes place.

 Consider the fluctuations of the surface elevation, $\delta h({\bf{x}}, t)$, which depends on space, i.e.
     on the horizontal coordinates
    $\bf{x}$ (boldface are for vectors in the Euclidean geometrical space) and on time $t$.
From the data, the only scaling parameters are the length
$\lambda$ and the time $\tau$. Given $\epsilon$, the fluctuations
$\delta h({\bf{x}}, t)$ may then be written as
\begin{equation}
     \delta h({\bf{x}}, t) = \lambda \delta H(\frac{{\bf{x}}}
     {\lambda}, \frac{t}{\tau})
    \mathrm{.}
\label{eq:scalingelevation}
\end{equation}
Therefore we expect that  $\delta h({\bf{x}}, t)$ is of order
$\lambda$, and changes with respect to space and time with the
typical scales $\lambda$, and $\tau$, eventually times factors
depending on $\epsilon$ only. More precisely we expect that the
horizontal dependence is scaled with
  $\lambda$ without factor depending on $\epsilon$, at least for $\epsilon$ small, because the typical
   wave-length of the unstable fluctuations
  is of order $\lambda$. This is in qualitative agreement with the observation that, the larger the
   wind speed the larger is the average wave-length of the wave system, as reported for instance in \cite{gelci}.
   Note that, if this average wave-length were much less than $\lambda$ (i.e. if the surface was flat at the scale
    $\lambda$)
   , this wave system would be unstable against fluctuations
   at wavelength of order $\lambda$, which is impossible if the wave system has reached a statistically steady state.
   Moreover this kind of assumption is fully consistent with the Pierson-Moscowitz spectrum where the wave period is of
   order $\tau$
    although the amplitude is multiplied by a small numerical prefactor of order $10^{-2}$ explained below by a
    non-trivial dependence of the average wave-amplitude on $\epsilon$ \cite{comment}. Indeed the wave-height should
    scale like $f(\epsilon) \lambda$, where the $f(.)$ function tends to zero as its
  argument
   tends to zero ( if $f(0)$ would not be
   zero, the water-wave would keep a finite amplitude in
   absence of wind,
    which
    is against common sense, because there are no waves if no instability feeds the wave system).
    Therefore
     the "scaled" stochastic function $\delta H({\bf{X}}, T)$, with
    capital letters for scaled quantities, is proportional to the reducing factor $f(\epsilon)$.
    We shall prove that $f(\epsilon)$ is logarithmic with
    respect to $\epsilon$. Assuming invariance under translation in space and
      time
      ,  the pair correlation $ S_{\delta H}(\textbf{X},T;\textbf{Y},T+T') = <\delta H({\bf{X}}, T) \delta H({\bf{X}}
      + {\bf{Y}}, T + T')>$ is a function of $({\bf{Y}},T')$ and of $\epsilon$ only. This does not mean that it is
      isotropic, as one expects the direction of the wind to induce such anistropy.

      Let us now comment on
      the neglect of the viscosity of air and of water, and of the surface tension $\sigma$.
      The latter
introduces a new length scale, the capillary length $\lambda_c =
 \sqrt{\frac{\sigma}{g \rho_{w}}}$. Although surface tension of real sea-water is a highly variable quantity, the
 capillary length is somewhere between one centimeter and one millimeter, far smaller than the length scales we are
 concerned with. This does not mean however that its effects are negligible: surface tension could regularize the
 instabilities at the shortest scales, which otherwise could lead to the development of cone-like singularities on the
  free surface \cite{IHP}, while white caps are observed instead. Here we are interested on large scale phenomena only,
  then capillarity effects are not
  considered,
   and the capillary length scale will
   be taken as zero.
   Here we consider high wind velocity regime, more precisely the
   regime above the onset of wave-breaking \cite{windspeed} which occurs for  $U_{10}$ around $7m/s$, ( $U_{10}$ is the
   wind speed measured
    $10$ meters above the sea surface), where gravity becomes the only governing parameter of wind-wave
   interaction. At inferior velocities, surface tension may either dominate gravity everywhere or just make impossible
    the steepening leading to wave breaking \cite{L-H}.
Similar statements could be made abouth the effect of viscosity:
if the Reynolds number is large, viscosity is relevant
 at very small scales that can be taken as simply zero.

 Now we would like to discuss the possible relationship between
 the ideas introduced above and the one of cascade of energy. The idea of cascade of energy is prevalent in fully
 developed turbulence, where it is assumed that energy goes step by step from large to small scales, energy being
 injected at the largest scales and dissipated at the tiny Kolmogorov length set by viscosity. The application of this
 idea to wind-driven waves meets several difficulties. The first one is that no scale is fixed from outside by the
 geometry of the flow, therefore it is hard to claim that
  energy is "injected" at large scales only. The wind acts on each scale, as illustrated in the different models using
  an input source term that is linear or quasi-linear with respect to the wave-spectrum. One must note that
 the Kelvin-Helmholtz instability is effective from $\lambda$, defined above, down to the capillary length $\lambda_c$,
  the
  smaller length scales being the most unstable. In such a fully nonlinear
  problem, the process of direct cascade (from large to small scales)
  has no serious support. Furthermore the numerical solution of the equation of weak turbulence shows instead a drift
  of the peak spectrum towards larger and larger scales \cite{elodie}.

  The reverse statement, that is that wave energy is continuously injected at small scales and cascade up to large scales
   is even more
   meaningless: at large scales viscosity is negligible and there is no dissipation (at least if "large scale" is meant
    in its general acception) and the energy has no way to go to in a steady statistical state (this does not apply to
    an initial value problem without sustained energy input).  Moreover an inverse cascade is in complete contradiction
    with casual observation: wind waves tend to propagate along the wind direction, although, resulting from an inverse cascade, their spectrum should become isotropic after a few steps in this cascade. To conclude this
    discussion, a step by step cascade in any direction is not the dominant process for energy transfer across different length scales
    and cannot explain dissipation.

     Another remark may be also relevant: the air flow above the sea surface is highly
     turbulent and one expects the formation of a Prandtl logarithmic layer there. The scaling parameter of  such a
     Prandtl layer is a flux of horizontal momentum per unit area of the surface. Compared to our scaling via the wind
     speed, this introduces logarithmic corrections that are, most likely, hardly detectable. However this scaling via
     the flux of horizontal momentum, as compared to scaling via the wind speed, is not completely without consequences.
      Actually this flux should be the same in the water below the surface, with an  horizontal momentum of the same
      order of magnitude as the aerial flux. The flux of horizontal momentum scales  like $\rho u^2$, therefore the
      average horizontal speed underwater is of order of $U \sqrt{\epsilon}$ ($U$ wind
      speed), for example  a wind speed of $60$ km/h should generate an underwater current
       of about $1.9$ km/h. This prediction fully agrees with the observations, $u_{sea}/U_{10}=0.032$ \cite{Hsu}, and the common "3 per cent rule", i.e.
        that the surface
      current is approximately $3$ per cent of the wind speed.

Having excluded dissipation by a  multi-step cascade of energy as the dominant mechanism of
dissipation, we are led to look at another mechanism, wave-breaking. This is known to be a strongly dissipative and fully
 nonlinear process, Therefore it could look impossible a priori to estimate the power lost per unit area by wave-breaking,
given for instance the pair correlation of the fluctuations of
height. This pair correlation assumes implicitely a smooth wave
propagation, namely a singly defined $h(x,y,t)$ and a non
self-crossing surface, although wave-breaking is a complex
nonlinear process, requiring to account for physical effects like
surface tension and viscosity. Nevertheless wave-breaking of
dominant waves (whose wavelengths are near the peak of the
spectrum) should be a relatively rare event, as we shall explain.
Let us precise that we may also include in the breaking process
 the micro-breakers, or small white caps, whose
density increases with the the wind speed, and saturates at
$U_{10}$ around 20m/s. In this regime, each wave of any wavelength
has a white cap on its crest.  Nevertheless,  in the stationary
state, the
 crest length of breaking waves is a small fraction of the total crest length.
Those white caps may replace the formation of conical or
wedge-like singularity \cite{IHP} of the inviscid equations. Note
that white caps appear also after the breaking of a very long
wave. Here we focus on the very large energy dissipation process
due to wave-breaking of all waves.

  Let us estimate the probability of such a process. In a single
wave-breaking event, the lost energy $W_{br}$ scales as $  \rho_w
U^2 \lambda^2 \sigma_{\delta h}$, i.e. the energy corresponding to
a "typical" volume, $\lambda^2 \sigma_{\delta h}$ of a wave having
an horizontal surface $\lambda^2$, a height $\sigma_{\delta h}$,
and propagating at velocity $U$ (which is around the velocity
reached by long waves, the peak waves, before breaking). By
comparison, the kinetic energy of the wind which occupies the same
volume, is smaller by a factor $\epsilon$. The balance of energy
may be obtained in the steady regime when the frequency of
occurence of weave-breaking is of order $\epsilon$, since in this
case the power lost per unit area is of the same order, $\epsilon
W_{br}/\lambda^2 \tau$.

    The result of these considerations is that, at a given time, the area covered by
     wave-breaking events is of order $\epsilon$ times the total area, independently on the wind speed. Note that this is
     not the proportion of area covered by foam, much bigger than the area covered by
      wave-breaking events in high wind conditions, because of the spreading of the foam by the wind, and because
      of the white caps.

Let us now estimate the $\epsilon$-dependence of the magnitude of
wave fluctuations. Given the statistics of the wave fluctuations,
and because wave-breaking is a rare event, it depends on the
probability of large fluctuations. Those large fluctuations are
rare because of the smallness of $\epsilon$. Said otherwise, the
scaling laws for the
 magnitude of the fluctuations have to be amended in order to take into account this smallness: otherwise, if the
 magnitude of the fluctuations was such that the typical wave length and the typical wave height were both $\lambda$,
  the probability of occurence of wave breaking would be of order one too, although we argued it is of order $\epsilon$.
   Therefore the amplitude of the wave should be small compared to $\lambda$, to make exceptional the nonlinear evolution
   towards wave-breaking.  This implies that, predominantly, the wave system is described by the linear approximation of
    the wave equations. Therefore, at a given location, and according to an idea that seems to have been stated first by
     Planck for the classical (Rayleigh) part of the black-body spectrum, the fluctuations of the free surface are
     predominantly Gaussian because they are made of a {\emph{linear}} superposition of waves
     with a continuous distribution of frequencies.  The most interesting quantity from the point of view of wave-breaking is not so
     much the amplitude of the wave, but its slope. Being given by a linear transformation of the amplitude it has also
     a Gaussian distribution. Therefore the derivative $\frac{\partial h}{\partial x} = h_{,x}$, namely
      the gradient of the height along $x$, the wind direction, has the  probability
      distribution
\begin{equation}
     {\mathcal{P}}(h_{,x}) =
     \frac{1}{(2\pi)^{1/2} \sigma_{x}} e^{-\frac{h_{,x}^2}{2\sigma_x^2}}
    \mathrm{,}
\label{eq:gauss}
\end{equation}

where $\sigma_x ^2$ is the variance of the slope along the $x$
direction.

 Wave-breaking is a nonlinear phenomenon
where the free surface becomes first vertical and then overturns.
Before this happens, the slope has to reach values of order one.
Indeed such a phenomenon is not described by the linear
approximation for wave propagation, since it assumes the height to
be a single valued smooth function of the horizontal coordinates.
Nevertheless we may assume that, before the wave locally
overturns,  its slope gets finite (non-small) values that are at
the border of applicability of the linear approximation
\cite{IHP}. The probability of overturning may be approximated as
the tail area of the density ${\mathcal{P}}(h_{,x})$, assuming
that the waves propagate predominantly in the wind direction),

     \begin{equation}
     {\mathcal{P}}_{br} =
     2\int_a^{\infty} {\mathcal{P}}(h_{,x}) d(h_{,x})= 2
      {\mathrm{erfc}}(\frac{a}{\sqrt{2}\sigma_x})
    \mathrm{,}
\label{eq:probabr}
\end{equation}
where $a$ is a parameter of order unity defining the limit slope
(above which the wave breaking occurs with a finite probability), and ${\mathrm{erfc}}(z)=
\frac{z}{\sqrt{\pi}}\int_z^{\infty} e^{-\varsigma^2} d\varsigma$.
In the limit of large $z$,  $
 {\mathrm{erfc}}(z)\simeq\frac{exp(-{z^2})}{z\sqrt{\pi}}$ . According to the
arguments presented before the probability ${\mathcal{P}}_{br}$
must be of order $\epsilon$ to ensure the balance between energy
input and dissipation by wave-breaking. In the limit $\epsilon <<
\sigma_x << a$, this gives
\begin{equation}
     \sigma_x \sim \frac
     {a} {\sqrt{\ln(\epsilon ^2)}}
    \mathrm{,}
\label{eq:sigma}
\end{equation}
This is of order $0.27$ for the case of wind-waves, if one takes
$a=1$.

This behavior is quite weak, in agreement with the observation
that the slope is small on average, but not very small
\cite{slope}. In this respect the density of breaking events is
far more sensitive to the smallness of $\epsilon$ than the wave
amplitude itself. Indeed it seems difficult to change $\epsilon$,
so the prediction of an $\epsilon$ dependence of the probability
of wave-breaking is hard to test. However one may think to a value
of $\epsilon$ close to 1, with water and oil for instance (E.J.
Wesfreid private communication), which would then yield a wave
system where wave breaking is almost everywhere, a clearcut prediction of the present theory.

Let us sketch now a more quantitative approach. As has been shown
since a rather long time \cite{hasselman}, small non
 linearities yield a kinetic equation for wave turbulence. In this theory the interaction between waves of various
 wavenumber and frequencies yields a kind of Boltzmann-like theory.
  In the Hasselmann-expansion,  written with respect to the elevation amplitude as small
  parameter, the nonlinear interaction between the waves appear as a serie involving successively four, six,
  etc..wave-interactions \cite{interactions}. Actually the
  relation (\ref{eq:sigma}) proves that the ratio
  between the six and four wave-interaction terms is always small.
  Therefore using our scalings, and the small slope variance as
  small parameter, we infer that formally the Hasselmann-expansion
  can be continued order by order (even for non small waves), at the price of
  fastly growing
 complexity. There is no reason that this expansion becomes ill-defined at any finite order. This does not mean however
  that, even by including the kinetic terms at all orders, all the physics is captured. This is the well-known phenomenon
  of expansion beyond all orders  \cite{byondallorder}. Indeed the occurence of wave breaking depends on effects transcendentally small with respect to the expansion
   parameter, here $(\ln\epsilon^2)^{-1/2} $.  The "standard" way of getting such transcendentally small
    terms is by looking at the general large order term in the expansion, getting its leading order part and then summing
    the largest terms of each order, something perhaps doable, but surely very cumbersome
    in the present case.

 Let us now include in a qualitative sense the dissipation due to
wave-breaking, in the light of the above estimations summarized in
equations (\ref{eq:probabr})-(\ref{eq:sigma}). Using the scales
variables, the steady state spectrum $N(\textbf{K})$ of the
surface elevation, which is the spatial Fourier transform of the
single time correlation function $S_{\delta
H}(\textbf{X},T;\textbf{X}+\textbf{Y},T))$, writes
\begin{equation}
{\mathcal{N}}({\bf{k}}) = \frac{1}{(2\pi)^2} \int
{\mathrm{d}}{\bf{Y}} e^{i{\bf{Y}}\cdot{\bf{K}}} S_{\delta H}(
{\bf{Y}}) \mathrm{.} \label{eq:spectrum}
\end{equation}

It obeys the stationary Hasselmann equation \cite{hasselman},
\begin{equation}
S_{nl}[{\mathcal{N}}({\bf{K}})]+S_{in}[{\mathcal{N}}({\bf{K}})]+S_{diss}[{\mathcal{N}}({\bf{K}})]=0
    \mathrm{,}
\label{eq:eqNk}
\end{equation}
 where the term $S_{nl}$ represents the exchange by
nonlinear interaction  between waves, $S_{in}$ the
input by the wind, and $S_{diss}$ for the dissipation by
whitecaps and wave-breaking of dominant waves.

 The term $S_{nl}$
describing the transfer across the spectrum of
fluctuations due to four wave interactions writes

\begin{eqnarray}
S_{nl}[{\mathcal{N}}({\bf{K}})] &=& \int  {\mathrm{d}}{\bf{K}}_1
{\mathrm{d}}{\bf{K}}_2  {\mathrm{d}}{\bf{K}}_3 |T_{0123}|^2 \delta
({\bf{K}}+ {\bf{K}}_1 - {\bf{K}}_2  - {\bf{K}}_3)
\delta (\Omega+ \Omega_1 - \Omega_2  - \Omega_3) \times \nonumber \\
 & & \quad {\mathcal{N}}({\bf{K}}_1){\mathcal{N}}({\bf{K}}_2){\mathcal{N}}({\bf{K}}_3){\mathcal{N}}({\bf{K}})
\left(\frac{1}{{\mathcal{N}}({\bf{K}}_1)} +
\frac{1}{{\mathcal{N}}({\bf{K}})}  -
\frac{1}{{\mathcal{N}}({\bf{K}}_2)}  -
\frac{1}{{\mathcal{N}}({\bf{K}}_3)}  \right) \mathrm{,}
\label{eq:kineticeq}
\end{eqnarray}

where $\delta(.)$ is the Dirac function. The transition matrix
$T_{0123}=T({\bf{K}}, {\bf{K}}_1, {\bf{K}}_2, {\bf{K}}_3)$ has a
rather complex explicit form \cite{webb},
 and is equal to $K^{5/2}$ times a numerical function of the ratios $K/K_i$, with $i=1,2,3$, and of the
 angles between the four vectors $({\bf{K}}, {\bf{K}}_1, {\bf{K}}_2, {\bf{K}}_3)$.  Moreover $\Omega= \tau \sqrt{gk}$
 , and
 $\Omega_i= \tau \sqrt{gk_i}$ .
The input from the wind  is just $\epsilon
{\mathcal{N}}({\bf{K}}) K_X $, where $K_X = k_x$ is the Cartesian
component of $\bf(k)$ along the wind. This simple approximation could be refined by multiplication by a function
 $G({\bf{K}})$ representing in dimensionless notations the detailed dependence of the input as a function of the
  wave number of the waves. We just take $G =1$ to make the exposition simpler.

 An essential point  of our
formulation is that, in our dimensionless variables this term is
small, proportional to $\epsilon$, this smallness being a
consequence of the scaling laws, not assumed from the beginning.
Moreover this term is proportional to $ k_x$ to represent the
angular dependence of the rate of growth of the Kelvin-Helmholtz
instability, and it is also proportional to the intensity of the
fluctuations because this is a linear instability.

The energy loss by wave-breaking is equal to $
b'{\mathcal{P}}_{br} {\mathcal{N}}({\bf{K}})$ because it is
proportional to the small probability of this process, and it
should be proportional to the spectrum itself. The numerical
factor $b'$ is  of order $1$, as discussed below. Using the
expression (\ref{eq:probabr}), with $\sigma_x ^2= \int
{\mathrm{d}}{\bf{K}} K_X^2 {\mathcal{N}}({\bf{K}})$, the balance
equation (\ref{eq:eqNk}) leads to the following integral equation
for the surface elevation spectrum in steady situations,
\begin{equation}
{\mathcal{S}}_{nl} [{\mathcal{N}}({\bf{k}})]+ \left(\epsilon  K_X
- \frac{1}{ab }\sigma_x
 e^{-a^2/2 \sigma_x ^2} \right)  {\mathcal{N}}({\bf{k}})=0
 \mathrm{,}
\label{eq:kineticeqfull}
\end{equation}

where the constant $b=\frac{\sqrt{\pi /2}}{b'}$ is the duration of
the breaking process in units of $\tau$. There are two unknown
parameters, $a$ and $b$. Contrary to similar equations in the
literature \cite{zakharov}, the breaking-wave loss term is
{\emph{not}} proportional to a power of the amplitude
  of the fluctuations, it depends transcendentally on this amplitude, because wave-breaking is not found at any
  algebraic
   order in the amplitude expansion of the kinetic equations.  Although $g$ and $U$ have been scaled out, the small
   dimensionless parameter $\epsilon$ remains, in
particular because of the term depending transcendentally on the
unknown amplitude.  This integral equation yields the scaling laws
for the amplitude of the fluctuations by noticing that the first
term has "conservation laws". This is a familiar property of
Boltzmann-type equations, which writes
  $\int{\mathrm{d}}{\bf{K}} F({\bf{K}}) {\mathcal{S}}_{nl}  [{\mathcal{N}}({\bf{K}})]  = 0$ with $F({\bf{K}})
   = 1\mathrm{,}$ ${\bf{K}}$ or $ \Omega(K) \mathrm{.}$ Therefore one finds three relations to be satisfied by
   the steady spectrum :
\begin{equation}
\int{\mathrm{d}}{\bf{k}} F({\bf{k}}) \left(\epsilon  K_X -
\frac{1}{ab } \sigma_x
 e^{-a^2/2 \sigma_x ^2} \right)  {\mathcal{N}}({\bf{k}}) = 0
 \mathrm{,}
\label{eq:kineticeqconserv}
\end{equation}
This yields the same scaling relation as derived before, if one
assumes that the integration over $\bf{K}$ and the multiplication
by $F(.)$ change the scaling in the same way in the two terms of
equation (\ref{eq:kineticeqconserv}), the one representing the
input by the instability (proportional to $\epsilon  k_x U$) and
the one representing the loss by wave-breaking, proportional to
the exponential. Notice too that, thanks to the prefactor $
\sigma_x$ in the dissipation part, the dissipation becomes more
efficient as the amplitude of the spectrum grows. This makes
likely impossible a steady (unphysical) spectrum spreading without
decay to infinitely large wave numbers.

 To summarize, we have introduced the idea of a physical small parameter in the wind-sea interaction, the ratio of the
  mass densities of the two fluids. Thanks to that, we have been able to derive scaling laws for the physically
  observable quantities like the mean square slope of the sea surface. This gives a basis for
the derivation of an Hasselmann-type equation for the steady state
spectrum, that would be valid not only for the case of wealkly
nonlinear turbulence, where it is generally adressed, but even for
storms with very large wind velocities. We propose a fully
explicit mathematical model,
  equation (\ref{eq:kineticeqfull}), for the steady spectrum
  of surface elevation perturbed by a constant wind. This model is valid for large wind speeds, where dissipation is
  mostly due to wave-breaking, although at low speeds the
  wave-breaking is stopped (before the overhanging of the sea surface) by capillarity effects.
  Indeed all this remains to be confronted with experimental results alhough the solution of
  the spectral equation remains to be studied in details in the limit $\epsilon$ small.
{}

\thebibliography{99}
\bibitem{PM} Pierson W.J. and  Moskowitz L. " A proposed spectral form for fully developed wind seas based on the
similiarity theory of S.A. Kitaigorodskii" , J. Geophys. Res.
{\bf{69}} (1964) 5181-5190 .
 \bibitem
 {gelci} Gelci R., Cazal\'e H. and Vassal J. "Pr\'evision de la houle. La
m\'ethode des densit\'es spectroangulaires", Bulletin
d'information du Comit\'e d'Oc\'eanographie et d'Etude des
C\^{o}tes {\bf{9}} (1957) 416 - 435.
\bibitem{hasselman} Hasselmann K.  "On the nonlinear energy transfer in a gravity-wave spectrum, part I: general
 theory",  J. Fluid Mech. {\bf{12}} (1962) 481-500 ; "On the spectral dissipation of ocean waves due to white
 capping.",
Boundary-Layer Meteorol. {\bf{6}} (1974) 107-127;  Hasselmann S.,
 Hasselmann K. , Allender J. H. and Barnett J. P. "Computations and Parametrizations of the nonlinear energy transfer
  in a gravity wave spectrum. Part II", J. of Physical
Oceanography {\bf{15}} (1985) 1378-1391.
\bibitem {YPscaling} Pomeau Y.  " Asymptotic time behaviour of nonlinear classical field
equations",  Nonlinearity  {\bf{5}} (1992) 707-720.
\bibitem{comment} The classical Pierson-Moscowitz frequency spectrum writes ${\mathcal{N}}(\omega)
= \frac{\alpha g^2}{\omega^5} exp\left[ - \beta (\omega
\tau)^{-4}\right]$ where $\alpha$ is a small numerical constant,
of order $8. 10^{-3}$, while the constant $\beta$ is about $0.74$.
The amplitude of the waves, compared to $\lambda$, is measured by
$\alpha$. We explain the smallness of $\alpha$ by the smallness of
the input
 from the wind, although $\beta$ compares the average wave frequency $1/\tau$, and is therefore of order unity.
\bibitem{IHP} Pomeau Y. and Le Berre M. "Topics in the theory of wave-breaking", submitted.
\bibitem{windspeed} Amorocho J. and DeVries J.J. "A new
evaluation of the wind stress coefficient over water surfaces"
 J. Geophys. Res. C: Oceans Atmos. \textbf{85} (1980)
433-442.
\bibitem{L-H} Longuet-Higgins M. S. "On Wave-breaking and
the equilibrium spectrum of wind-generated waves", Proc. Roy. Soc.
{\bf{A  310}} (1969) 151-159; "On the form of the highest
progressive and standing waves in deep waters", Proc. Roy. Soc.
{\bf{A 331}} (1969) 445-546.
\bibitem{elodie} Benoit M. and Gagnaire-Renou E.  "Int\'{e}ractions vague-vague nonlin\'{e}aires et spectre d'
\'{e}quilibre pour les vagues de gravit\'{e} en grande profondeur
d'eau", irevue (2007) 18 \`{e}me congr\`{e}s fran?ais de
m\'{e}canique, S-19 Ondes et \'{e}coulements en surface libre.
\bibitem{Hsu} Hsu S.A.  "Estimating overwater friction
velocity and exponent of power-law wind profile from gust factor
during storms", Journ. of waterway, Port, Coastal and ocean eng. ,
{\bf{129}} (2003) 174-177.
\bibitem{slope} While it is difficult to compare our predictions with real systems, because we have assumed a steady state
in an infinite sea, we may mention the satellite observations
performed for unsteady wind-waves, averaged over the sea surface
of the Earth reported by Hu Y. et al.
 "Sea surface wind speed estimation from space-based lidar measurements", Atmos. Chem. Phys. Discuss.
 {\bf{8}} (2008)
 2771-2793.
 They found values of $\sigma_x$ around $0.23-0.3$, for wind
 velocity around $10m/s$ to $20 m/s$. This result approximately agrees with measurements done in tank experiments,
as reported by  Wu J. "Directional slope and curvature
distributions
 of wind waves", J. Fluid Mech.  \textbf{79} (1977)
463-480.
\bibitem{interactions} Three-waves interaction is not permitted by the resonance condition on the frequencies and wave-numbers, that
  reads $\omega(k_1) + \omega(k_2) = \omega(k_3)$ and the wave numbers ${\bf{k}}_1 + {\bf{k}}_2 = {\bf{k}}_3$, impossible
   to satisfy with $\omega(k) = \sqrt{gk}$, although the equivalent condition for the four wave interaction  $\omega(k_1)
   + \omega(k_2) = \omega(k_3) + \omega(k_4)$ and ${\bf{k}}_1 + {\bf{k}}_2 = {\bf{k}}_3 + {\bf{k}}_4$ can be satisfied.
       \bibitem{byondallorder} "Asymptotics beyond all orders" Ed. by Segur H., Tanveer S. and Levine H. , Nato ASI series,
        Series
        B Physics, Vol. 284 Plenum press, New York (1991), Proceedings of a Nato ARW at La Jolla CA
        (Jan.1991).

\bibitem{webb} Webb D. J. "Nonlinear transfer between sea waves", Deep-sea
res. {\bf{25}} (1978) 279-298.
\bibitem{zakharov} See for example Filipot J.-F., Ardhuin F. , Babanin A. and  Magne R. "Param\'{e}trage du d\'{e}ferlement
des vagues dans des mod\`{e}les spectraux: approches
semi-empirique et physique", X\'{e}me Journ\'{e}es Nationales
G\'{e}nie C\^{o}tier, (2008), Sophia-Antipolis, France. See also
the review paper by Badulin S.I., Pushkarev A.N., Resio D. and
Zakharov V.E. "self-similarity of wind-driven seas" Nonlinear
Processes in Geophysics {\bf{12}} (2005) 891-945.
  \endthebibliography
\end{document}